\documentclass[final,5p,times,twocolumn]{elsarticle} 

\usepackage{lineno,hyperref}
\modulolinenumbers[5]

\usepackage{siunitx}
\DeclareSIUnit\bit{bit}

\usepackage{graphicx}

\usepackage{caption}
\usepackage{subcaption}

\journal{Nuclear Instruments and Methods in Physics Research, Section A}

\begin{document}

\begin{frontmatter}

\title{Timing and Synchronization of the\\ DUNE Neutrino Detector}

\author[uob]{D.~ Cussans}
\author{on behalf of the DUNE collaboration.}

\address[uob]{University of Bristol, UK}


\begin{abstract}
	The DUNE neutrino experiment far detector has a fiducial mass of \SI{40}{\kilo \tonne}. The O(1 M) readout channels are distributed over the 4 x 10kt modules and need to be synchronized to O( \SI{10}{\nano \second} ) with a reliable, simple, affordable system. For the majority of channels a simple DC-balanced protocol is used, with clock and synchronization information encoded on the same fibre. The remaining channels use ``White Rabbit'' (IEEE-1588). Small scale tests show a timing jitter of \textless \SI{100}{\pico \second }. The DUNE timing system has been successfully prototyped at the ProtoDUNE-SP detector at CERN.   

\end{abstract}


\end{frontmatter}


\section{The DUNE Experiment}

The Deep Underground Neutrino Experiment (DUNE)\cite{ref:DUNE} will detect neutrinos generated 1,300km away in Fermilab, near Chicago. The far detector will consist of four modules, each with 10kt fiducial mass of liquid Argon, located 1.5km underground at the Sanford Underground Research Facility (SURF).

The energy deposited by the products of neutrino interactions will be detected in two ways: ionization drifted by an applied electric field to pick-up electrodes, forming a time projection chamber (TPC), and scintillation light.

In order to reconstruct events with the photon detection system readout channels must be synchronized to each other with a precision of nanoseconds. To correlate events in the far detector with neutrino generation at Fermilab the entire detector must be synchronized to GPS time to a precision of O( \SI{100}{\nano \second} ). 

The first module will be a “single phase” design where charge is drifted directly onto an anode plane consisting of a grid of wires. Single phase module(s) will use the synchronization protocol described here and tested at protoDUNE-SP. 

At least one additional subsequent module will be a “dual phase” design where charge is drifted upwards through the liquid into the gas phase where charge multiplication is used to amplify the signal. Dual phase modules will use the White Rabbit implementation of IEEE-1588\cite{ref:IEEE1588}.

\section{protocol}

A continuous stream of 8b/10b\cite{ref:8b10b} encoded serial data is transmitted over a 1000Base-Bx\cite{ref:ethernet} optical link from the timing master to timing “end points”.  In DUNE a 62.5MHz clock will be derived from a \SI{312}{\mega \bit / \second} data stream. In ProtoDUNE a 50MHz clock and \SI{250}{\mega \bit / \second} data stream were used. The data carries commands that are used to synchronize a 64-bit time-stamp counter in the end-point and propagate commands and synchronization messages. This allows the readout clocks, calibration signals, etc. to be synchronized throughout each module.

Each link from the timing master is bidirectional with data transmitted in both directions down a single fibre with different light wavelengths. Normally, data are only transmitted from the timing master to the endpoints. However, the data can be echoed by the end points and transmitted back to the timing master. This allows the timing master to measure the latency between timing master and end point and adjust the clock phase and time-stamp at the end point. This allows all end points to operate with the same time-stamp and clock phase. For links where a single fibre from the timing master is passively split to many end points only one end point at a time can transmit back to the timing master. The end points can also report their status back to the timing master. 

The protocol allows all endpoints to be addressed, groups of endpoints (partitions) or individual endpoints

\begin{figure*}
	\includegraphics[width=0.9\linewidth]{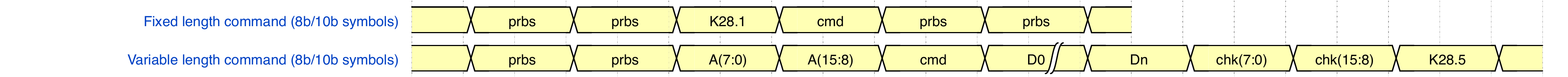}
	\caption{Example DUNE timing system messages. Fixed length and variable length}
	\label{fig:DTS_protocol}
\end{figure*}

Two types of messages are defined,
fixed length messages, which have fixed latency transmission from timing master to end-point, and variable length messages. Figure \ref{fig:DTS_protocol} shows examples of fixed and variable length messages.

The DUNE timing system protocol has been designed to be simple and robust. It allows the use of passive optical splitting, which reduces cost by allowing a single fibre to serve multiple end-points. It also allows passive optical combining which permits hot-swap redundant timing masters.

Because of its simplicity, the end-point circuitry can be easily implemented with an FPGA and few external components. The FPGA logic resources used are very much smaller than are needed for timing protocols that require the use of a micro-processor, such as IEEE-1588.

\section{Laboratory Tests}

\begin{figure*}
	\includegraphics[width=0.9\linewidth]{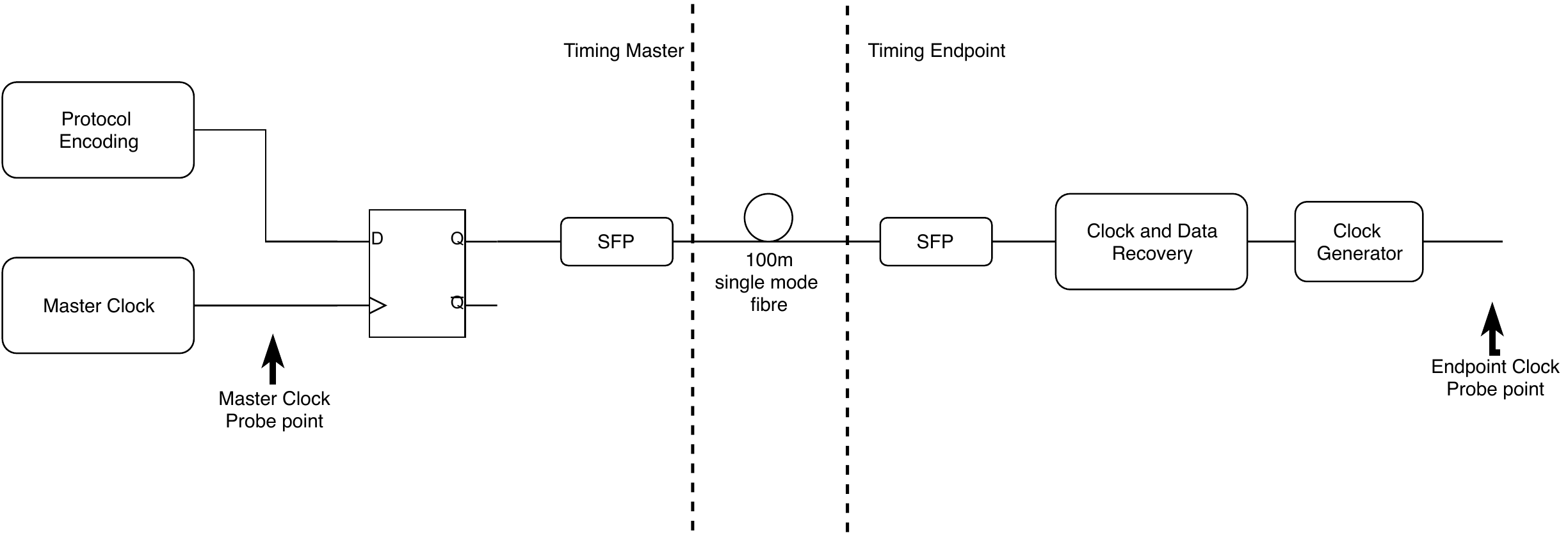}
	\caption{Block diagram of jitter performance test setup}
	\label{fig:jitter_setup}%
\end{figure*}

\begin{figure}
	\includegraphics[width=0.9\columnwidth]{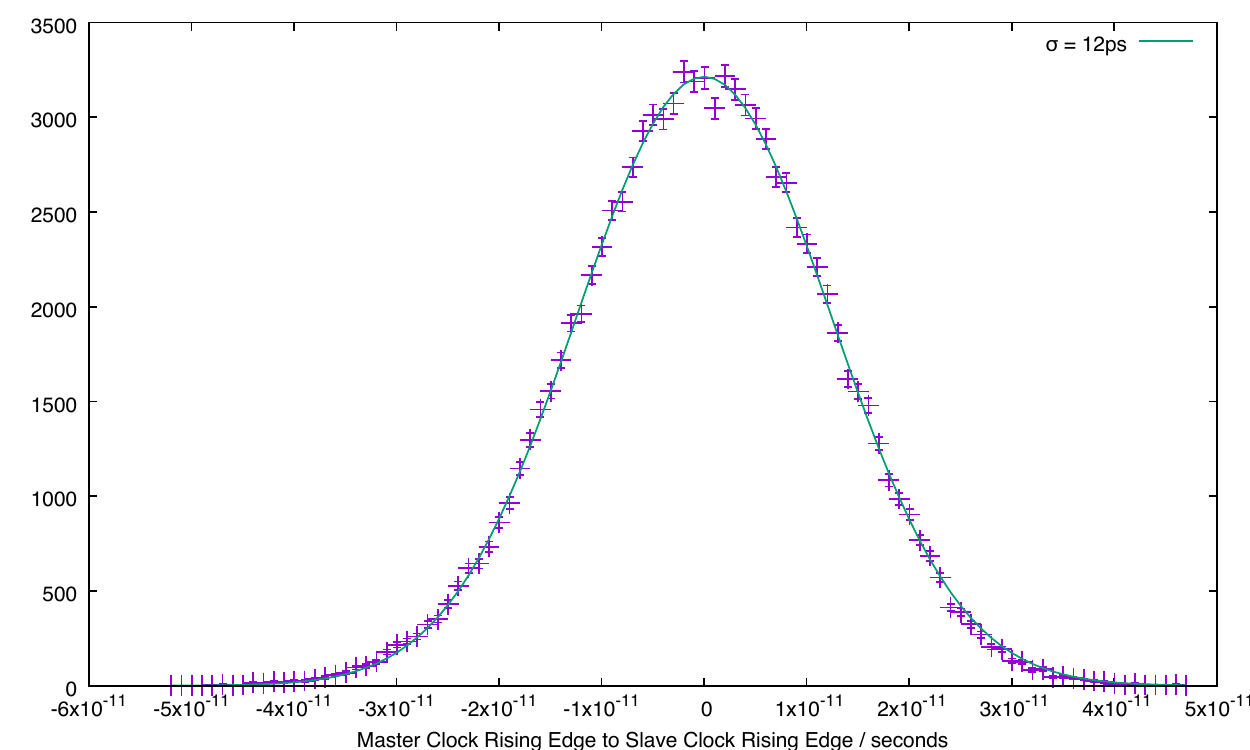}%
	\caption{Distribution of time difference between clock edges at timing master and endpoint }
	\label{fig:timing_jitter}
\end{figure}

Timing jitter performance was measured in the laboratory by connecting a prototype timing master with a prototype endpoint by optical fibre. The clock signals in the master and endpoint were compared. Timing jitter depends on the bandwidth programmed into the PLL clock generators used but is always less than \SI{100}{\pico \second} and typically less than \SI{20}{\pico \second}  . A block diagram of the apparatus used to test cycle-to-cycle timing jitter is shown in figure \ref{fig:jitter_setup}. The distribution of the time difference between clock edges at the timing master and timing endpoint is shown in figure \ref{fig:timing_jitter}. In other tests over 100 endpoints were synchronized.


\section{Beam Test (ProtoDUNE)}

\begin{figure}
	\centering
	\includegraphics[width=0.8\columnwidth]{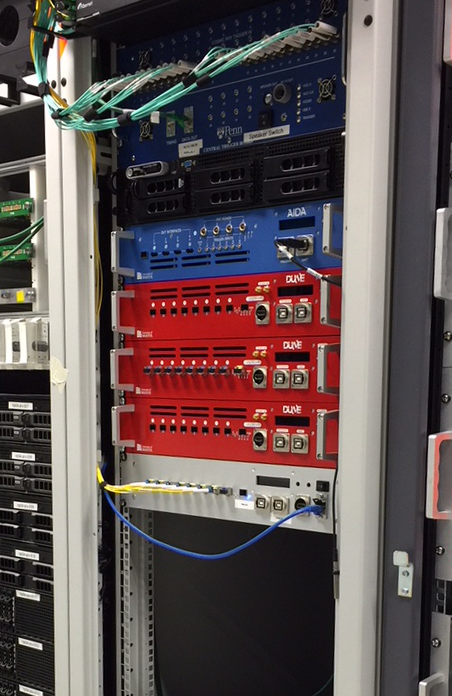}
	\caption{ProtoDUNE Timing System hardware installed at the CERN Neutrino Platform and waiting cabling. The AIDA-2020 TLU with blue front panel is above three active fanout units with red front panels.}
	\label{fig:pdts_hardware_installed}
\end{figure}

The DUNE timing system was used to synchronize the ProtoDUNE-SP experiment performed at the CERN Neutrino Platform. An AIDA-2020\cite{ref:AIDA2020} Trigger Logic Unit was used to accept accelerator synchronization signals and generate a timing data stream that was then fanned out to the TPC and photon detector readout systems. Figure \ref{fig:pdts_hardware_installed} shows the timing system hardware installed in ProtoDUNE.

As will be done in DUNE, active fanout units were followed by passive optical splitter/combiner units allowing each fibre from the master timing system to connect to multiple timing end-points.

Features such as measuring the master to endpoint latency were also tested.

\section{Implementation at DUNE}

\begin{figure*}
\includegraphics[width=0.6\linewidth]{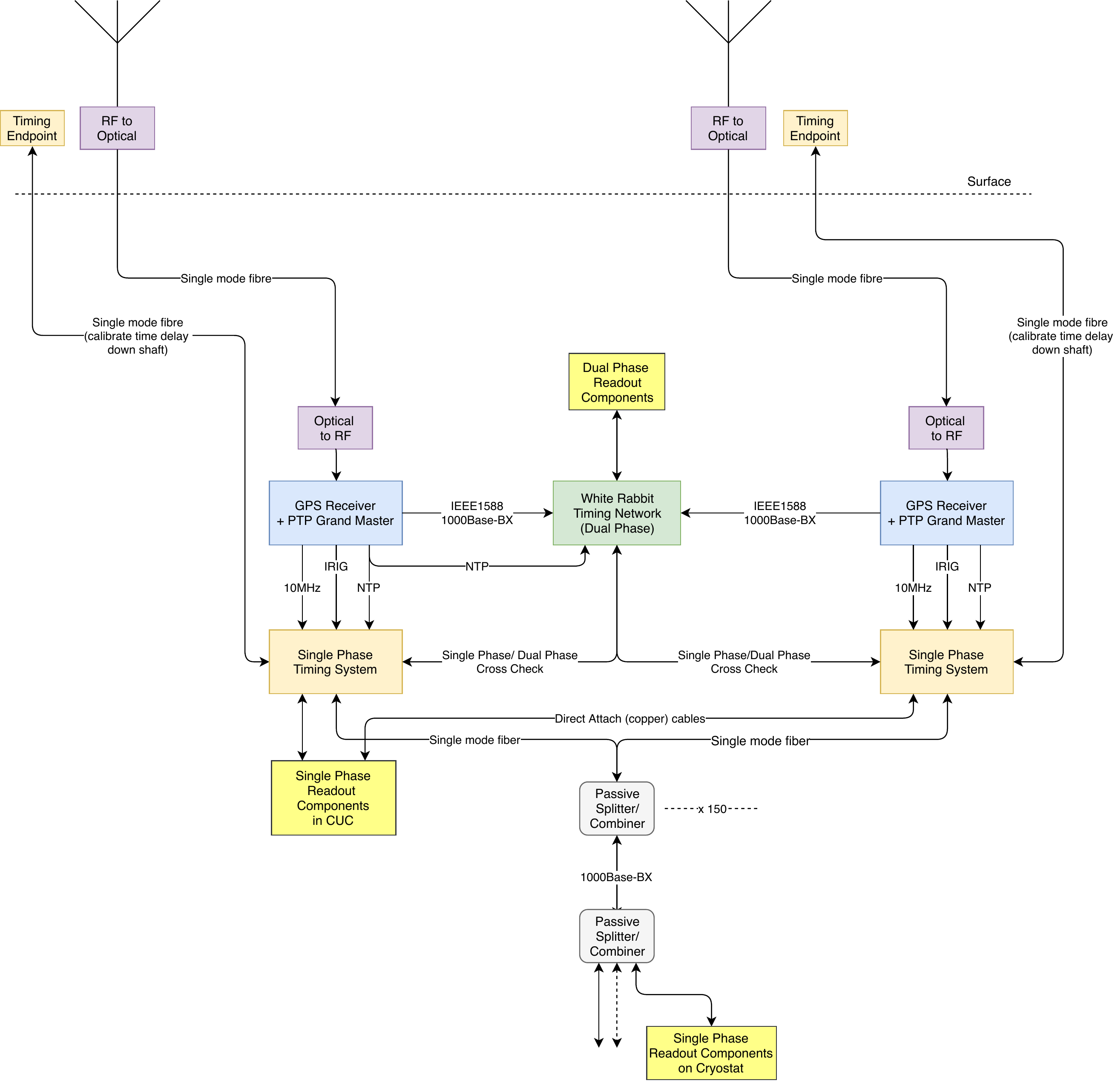}%
	\caption{Schematic view of the DUNE timing system.}
	\label{fig:DTS_overall}
\end{figure*}

\begin{figure*}%
	\includegraphics[width=0.6\linewidth]{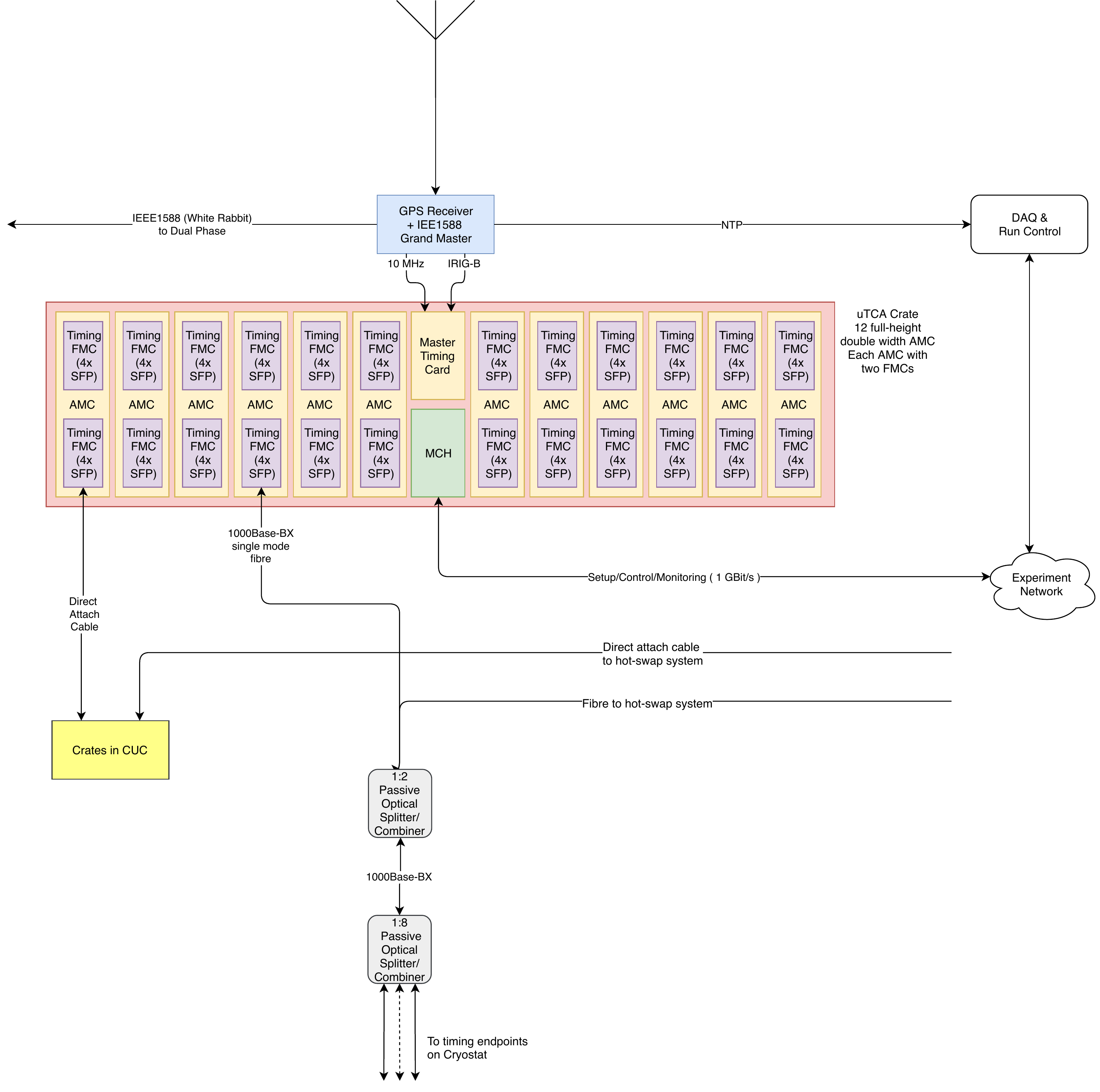}%
	\caption{Block diagram of timing system uTCA crate.}
	\label{fig:DTS_uTCA}
\end{figure*}

The DUNE timing system will be made largely out of commercially available components. MicroTCA crates will house commercial double width AMC boards each carrying two custom FMCs. In each crate the link to the GPS receivers will be done by a single custom AMC. Figure \ref{fig:DTS_overall} illustrates the overall layout of the DUNE timing system. Figure \ref{fig:DTS_uTCA} shows the arrangement of modules in each of the two microTCA crates.

The ability of the DUNE timing system protocol to be used with passive optical splitting and combining will be used to provide hot-swap redundancy to the system. The two completely independent master timing systems will allow firmware and software updates without interrupting system running. Two independent antennae will be used: One at the top each of the two access shafts to SURF. This, together with a rubidium atomic clock in the master timing system will enable failure of the link to GPS time to be detected.


\section*{References}

\bibliography{mybibfile}

\end{document}